\documentclass[prd,a4paper,nofootinbib,
showpacs, twocolumn,
]{revtex4}
\usepackage{graphicx}
\usepackage{amsfonts}
\usepackage{amssymb}
\def\eq#1{{eq.~(\ref{#1})}}

\def\etal{{\it et al.}}

\def\hbar{\hspace{0pt}\raisebox{1pt}{$-$} \hspace{-7pt} h}

\def\5{\overline 5}

\newcommand{\be}{\begin{equation}}
\newcommand{\ee}{\end{equation}}
\newcommand{\bea}{\begin{eqnarray}}
\newcommand{\eea}{\end{eqnarray}}
\newcommand{\nn}{\nonumber}

\begin{document}
\title[]{The little hierarchy problem\\
 for new physics just beyond the LHC
}
\date{\today}
\author{F.\ Bazzocchi$^{\dag\ddag}$}
\author{M.\ Fabbrichesi$^{\ddag}$}
\affiliation{$^{\ddag}$INFN, Sezione di Trieste}
\affiliation{$^{\dag}$SISSA, via Bonomea 265, 34136 Trieste, Italy}

\begin{abstract}
\noindent   We discuss  two possible extensions to the standard model in which  an inert singlet scalar state that only interacts with the Higgs boson is added together with some fermions. In one model the fermions provide  for a see-saw mechanism for the neutrino masses, in the other model for grand unification of the gauge couplings. Masses and interaction strengths are fixed by the  requirement of  controlling the finite one-loop  corrections to the Higgs boson mass thus addressing the little hierarchy problem.  The inert scalar could provide a viable dark matter candidate. Direct detection of this scalar singlet in  nuclear scattering experiments is possible with a cross section within reach of  future experiments.
\end{abstract}

\maketitle

{\bf 1.} If the absence of new states below the TeV scale~\cite{NPbounds} will be confirmed as the integrated luminosity of the LHC increases  in the next few years,   it will become unfortunately necessary to move  the  scale at which to expect new physics  outside the reach of the  experiments. Such an higher scale is somewhat in agreement with what already found at LEP  where a  cutoff scale for higher order operators encoding new physics is found to be larger than 5 TeV~\cite{Barbieri:2000gf}. Recent fits of supersymmetric models~\cite{susy} also indicate that the masses of the new particles may be just beyond the LHC reach and between 5 and 10 TeV.

The presence of new physics above 5 TeV rises the problem of the little hierarchy:     For the Higgs boson mass~\cite{Higgsmass}  (and the electroweak (EW) vacuum expectation value)  to be in the 100 GeV range---that is, roughly between one and two orders of magnitude  smaller than the new physics scale---renormalization effects must at least partially cancel out in order to prevent the Higgs boson mass from shifting to the higher energy scale. 

One may  implement such a cancellation by an appropriated choice of the Higgs boson bare mass but this would imply a fine-tuning of such a counter-term in which low and high-energy degrees of freedom are mixed. A more natural choice requires that the cancellation occurs at the higher scale and either comes from a symmetry or is an accident in which the various terms conspire to cancel against each other. In the latter case,  the cancellation is best thought as the effect of a  dynamical mechanism, at work at the high energy scale, which  arises from new physics that we do not know. The built-in fine-tuning of such a conspiracy (the same we would have at the level of the counter-terms) is of the order of the ratio of the two energy scales, in our case of about 10\%.

In what follows, we come back to the little hierarchy problem by following  the  empirical approach and discuss two possible scenarios for new physics: a representative see-saw model for neutrino masses and a grand unification model. In both of them, the addition of new states would shift the Higgs boson mass to the new scale unless we  balance the new contributions to prevent large one-loop renormalizations.  The identification of what states (their  masses and couplings to the Higgs boson)  must be present for such a balancing act to occur  provides the heuristic power of the little hierarchy problem.

While many possible new states can be added to prevent large corrections to the Higgs boson mass, the simplest choice consists in including just an  inert scalar state~\cite{inerts}, that is, a scalar particle only interacting with the Higgs boson (and gravity) thus transforming as the  singlet representation of the EW gauge  group $SU(2)\times U(1)$ (and similarly not charged under the color group) and  which acquire no vacuum expectation value. Such a choice minimizes unwanted effects on EW radiative corrections and other physics well described by the standard model (SM).

If in addition we impose a $Z_2$ symmetry under which  the  inert scalar is odd and all the SM fields are even,  the new state will couple  to the SM Higgs doublet only through quartic interactions in the scalar potential. By construction, we only look for solutions with  vanishing vacuum expectation value, thus $Z_2$ is unbroken and after EW symmetry breaking the singlet state can, as we shall discuss, potentially be a viable  cold dark matter (DM) candidate. 

The little hierarchy problem is often discussed in terms of the quadratic divergence arising in the mass term of the Higgs boson in a momentum dependent regularization (or, equivalently, in a pole in $d=2$ dimensions in dimensional regularization). 
In the past they have been cancelled either by assuming a symmetry (usually, supersymmetry) or by assuming that the Veltman condition~\cite{veltman} is satisfied, namely that the new sector couples to the SM Higgs boson just so as to make  the one-loop quadratic divergences to the SM Higgs boson mass vanish (see \cite{kundu} for various applications of this idea).  These divergent terms are a different and independent problem from the one  discussed here which only depends on  integrating out the heavy modes in the low-energy effective theory.  The terms we  worry about are  finite terms similar to  those arising in a supersymmetric theory with soft mass terms where the quadratic divergencies are  cancelled while, after integrating out the heavy states, there are finite terms   whose contribution shifts the values of the Higgs boson mass. 


\vskip 0.5cm
{\bf 2.} The first  SM extension we consider is a representative see-saw model~\cite{seesaw} for the neutrino masses. Three right-handed neutrinos $N_i$ are added. The  lagrangian of the model  is given by the kinetic and Yukawa terms of the SM with  the addition of the neutrino Yukawa terms:
\be
\mathcal{L}=\mathcal{L}_{Y_{SM}}+ y_{ij}^\nu \bar{N}_i \tilde{H}^\dag L_j+ \frac{1}{2} M_{N_i} N_i N_i
\,.
\ee
We  work in the  basis in which the right-handed neutrino mass matrix is real and diagonal.   
 
We compute the one-loop  finite contributions to the Higgs boson mass using dimensional regularization with renormalization scale $\mu$.
 The SM particle contributions are negligible. To compute the one-loop contribution arising from the right-handed  neutrinos we rotate the Yukawa couplings $y^\nu_{ij}$ into the basis in which the neutrino mass matrix, defined as
\be
m_\nu= -y^{\nu T}\cdot \frac{1}{M_N}\cdot y^\nu  v_W^2\,,
\ee
is diagonal.  According to the  Casas-Ibarra parametrization~\cite{Casas:2001sr}
 we have that
\be
\hat{y}^\nu_{ij}=  (y^\nu U)_{ij} =M_{N_i}^{1/2} R^\dag_{ij} \hat{m}^{1/2}_{\nu_j}\,,
\ee
where $\hat{m}_{\nu}$ is the light neutrino diagonal mass matrix and $R$ an arbitrary orthogonal complex matrix.

In the traditional see-saw model the Yukawa couplings are of order one and the masses $M_{N_i}$ very large and close to the GUT scale. If the Yukawa couplings are taken to be small, the $M_{N_i}$ can be  accordingly lighter.

Taking into account the one-loop contribution, and assuming  right-handed neutrino degeneracy  as well as  $R$ real, the Higgs boson mass receives a shift given by
\be 
\label{finite}
 \frac{1}{16 \pi^2} \frac{M^3_{N}}{v_W^2}  \sum m_\nu \left(\frac{3}{2} - \log\frac{M^2_{N}}{\mu^2}\right) \, ,
\ee
being $\mu$ the matching scale that in this case we identify with $M_N$. 
The sum of the neutrino masses, the term $\sum m_\nu$ in \eq{finite}, has a lower bound of about $ 0.055$ eV~\cite{x}, which corresponds to a  normal neutrino mass  hierarchy with vanishing lightest mass. On the other hand, cosmological constraints set  an  upper bound   on $\sum m_\nu$  that, even if model dependent,  is always $\leq 0.44$ eV~\cite{GonzalezGarcia:2010un}. 

Because of the smallness of the neutrino mass term, as long as the new states have masses up to around  $10^4$ TeV, the shift in the Higgs boson mass is of the order of its mass and no hierarchy problem arises.  
Notice that the one-loop correction of  right-handed neutrinos with $M_N\sim 10^4$ TeV  gives rise to a correction  to the Higgs boson mass  of the order of
\be
  \left(\frac{\sqrt{M_{N} \sum m_\nu}}{v_W} \right) M_N \sim 2.5\, \mbox{TeV}\,,
\ee
for which also two-loop corrections are under control.

On the other hand, if the new fermion masses $M_{N_i} \simeq M_N$  are larger  than $10^4$ TeV, we do have a little hierarchy problem and must balance their one-loop contribution against some other contribution in order to keep the overall renormalization of the Higgs boson mass of the order of the weak scale.

To provide for such a contribution,  we add the simplest  state: an inert scalar particle $S$.
The scalar potential is given by
\bea
\label{pot}
V(H,S)&=&\mu_H^2 (H^\dag H)+ \mu^2_S S^2\nn\\
& +&\lambda_1 (H^\dag H)^2 +
\lambda_{2} S^4+ \lambda_{3} (H^\dag H ) S S \,.
\eea
Linear and trilinear terms for $S$ are absent due to the $Z_2$ symmetry mentioned above. 

Taking into account the one loop contribution induced by the  scalar state $S$  the overall shift to $\mu^2_H$, taking $\mu=M_S$ to minimize the logarithmic contributions to the matching,  becomes
\bea 
\label{finite2}
\delta\mu_H^2(M_S) &=& \frac{1}{16 \pi^2}\left[-\lambda_3 M_S^2 \nn  \right.\\
&-&  \left. \frac{M^3_{N}}{v_w^2}   \sum m_\nu \left(\log\frac{M^2_{N}}{M^2_S}-\frac{3}{2}\right)\right]\,.
\eea

We want the  correction in \eq{finite2} to be of the order of the Higgs boson mass itself. For simplicity, we can just impose that $\delta \mu^2_H=0$ and obtain
\be
\label{l3}
\lambda_3=\frac{3}{2} \left(\frac{M_{N}^3 \sum m_\nu}{M_S^2 v_W^2} \right)\left[1-\frac{4}{3} \log \frac{M_N^2}{M_S^2} \right]
\ee

In the region $ M_S\ll M_N$, $M_N>10^4$ TeV a cancellation is possible provided $\lambda_3$  is negative. $\lambda_3$ is bounded by 
\be
\label{bpot}
\lambda_3 \geq -2 \sqrt{\lambda_1 \lambda_2}\,,
\ee
to ensure the stability of the scalar potential at infinity.   The value of $\lambda_1$ is fixed by the value of the Higgs boson mass to be $\lambda_1\sim.13$. Eq.~({\ref{l3}) and the above  condition are satisfied for $M_S > 5$ TeV.

For $M_S$ around 10 TeV, $\lambda_3 \simeq 0.2$. As the value of $M_S$ comes close to that of $M_N$---and the logarithmic term becomes smaller---the value of $\lambda_3$ becomes positive and smaller; it is  of the order of $10^{-7}$ for $ M_S\simeq  M_N$.

The order of the  the one-loop contribution to the Higgs boson mass  is $\sqrt{\lambda_3} M_S$.
The two-loop contributions are under control as long as  this correction is $\sim 10$ TeV.
 
\vskip 0.5cm
{\bf 3.} The second  SM extension we discuss is one in which we introduce the minimal set of fermion providing gauge coupling unification. The same question has been addressed in the context of split-supersymmetry models~\cite{Giudice:2004tc}. The possible sets are given by $(Q+\bar{Q})+(D+\bar{D})$, two chiral couples of  left-handed fermions with quantum number identical to the left-handed quark doublet and right-handed down quark respectively, or by $(L+\bar{L})+V+G$, one chiral  couple of left-handed lepton-like fermion  and a wino-like as well as a  gluino-like fermion multiplets. We choose the first option as the minimal and representative set. They couple to the Higgs boson SM through the Yukawa lagrangian
\bea
\label{lag2a}
&& M_Q \bar{Q} Q+ M_D  \bar{D} D +k_1 \bar{Q} D H +  k_2  \bar{D} Q H^*,
\eea
and they give a shift to the Higgs boson mass equal to 
\be
 \frac{ |k|^2}{16 \pi^2}\left [   ( 3 M_Q^2- M_Q  M_D)-
 3  M_Q M_D\log \frac{ M_Q^2}{\mu^2}\right]\,,
\ee
with $|k|^2= |k_1|^2+|k_2|^2$ and  $M_Q \sim M_D$. We identify the  the matching scale  $\mu$ with $M_D$.

If the new fermions are lighter than 1 TeV there is no little hierarchy problem. On the other hand, if they  are heavier the problem exists and we introduce  a inert singlet scalar $S$ to  protect  the Higgs boson mass. Therefore,   we add the terms
\bea
\label{lag2}
&& k_{3_i}  \bar{Q} q_i S+   k_{4_i} d^c_{L_i} D  S+H.c.  - V(H,S)
\eea
to the   lagrangian \eq{lag2a}.
In \eq{lag2}  $V(H,S)$ coincides with  \eq{pot} with $S$ odd under an additional $Z_2$ symmetry. We have also  imposed for extra fermions to be odd under $Z_2$.  The total one-loop contribution to $\mu_H^2$ at the scale $\mu=M_S$  is   given by
\bea
\delta\mu^2_H(M_S)&=&
 \frac{1}{16 \pi^2}\left  [-\lambda_3 M_S^2 
+  |k|^2 ( 3 M_Q^2- M_Q  M_D)\right.\nn \\
 &-&  \left. \label{x}
 3 |k|^2 M_Q M_D\log\displaystyle{ \frac{ M_Q^2}{M_S^2}}\right]\,.
\eea

Let us consider the case in which all couplings are of order one.
As before, for simplicity, we just impose that $\delta \mu^2_H=0$.  Taking $ M_Q \sim M_S$ and  with   the singlet $S$   lightest $Z_2$-odd particle, this condition is satisfied by writing  $\lambda_3$ as a function of $|k|^2$:
 \be
 \lambda_3= |k|^2 \left(\frac{M_Q}{M_S}\right)^2 \left(2-3 \log \frac{M_Q^2}{M_S^2}
 \right)\sim  2 |k|^2 \,.
 \ee
Contrary to the previous example of the see-saw model, in this case it is always possible to find an appropriate value of $\lambda_3$ so as to  control the renormalization of the Higgs boson mass.

\vskip 0.5cm
{\bf 4.} We may ask whether in the two models considered the inert scalar $S$ is a viable    DM candidate.
It is a gauge singlet and therefore only interacts with the SM particles through  the Higgs boson $h$.  The point-like interaction $\lambda_{3}/2\, S S h h$  and the scattering mediated by $h$---both in the $s$ and  $t$ channels---contribute  to the cross section $S S \to hh$. The Higgs boson $h$ also  mediates  the scattering processes $S S \to f\bar{f}$, $S S \to W^+ W^-$, $S S \to Z Z$. 

It has been shown~\cite{Yaguna:2008hd} that a single inert singlet  that couples with the Higgs boson with a small coupling    is a realistic cold DM candidate with  a mass $\lesssim v_W$. In our case,  the singlet  may account for the correct relic density in the opposite regime where  its mass is $ \gg v_W$ and its coupling with the Higgs boson   relatively large.  In this case,  the  scattering amplitude is dominated by the pointlike  $S S \to hh$ vertex  which gives a contribution to the total cross section equal to
\be
\label{sigma}
\langle \sigma v \rangle  \simeq \frac{1}{16 \pi} \frac{\lambda_{3 }^2}{M_{S }^2}\,.
\ee

To estimate the viability of $S $ as DM candidate,  we make use of the approximated analytical  solution~\cite{Srednicki:1988ce}. The relic abundance $n_{\mathrm{DM}}$  is written as 
\be
\label{exp}
\frac{n_{\mathrm{DM}}}{s}= \sqrt{\frac{180}{\pi g_{*}}}\frac{1}{M_{pl} T_f \langle \sigma  v\rangle}\,,
\ee
where $M_{pl}$ is the Planck mass, $T_f$ is the freeze-out  temperature, which for our and similar  candidates  is given by $m_{S }/T_f\sim 26$. The constant $g_*= 106.75+1$  counts the number of  SM degrees of freedom in thermal equilibrium plus the additional degrees of freedom related to the singlets, $s$ is their total entropy density. Current data fit within the standard cosmological model  give a relic abundance with $\Omega_{\mathrm{DM}} h^2=0.112\pm 0.006$~\cite{PDG} which corresponds to 
\be
\label{value}
\frac{n_{\mathrm{DM}}}{s}= \frac{(0.40\pm 0.02)}{10^9 \, M_{S}/\mbox{GeV}}\,. \label{exp2}
\ee
By combining \eq{value} with \eq{sigma} we may write $\lambda_3$ as function of $M_S$ obtaining 
\be
\label{lambda3}
|\lambda_3|\simeq 0.44 \, \frac{M_S}{\mbox{TeV}}\,.
\ee

 In the first model we considered,   the condition \eq{lambda3} can only be satisfied in the case in which $ M_S\ll M_N$, as it is shown in Fig.~\ref{fig0}. For $ M_S\simeq M_N$, the smallness of the neutrino Yukawa couplings forces $\lambda_3$ to be very small  thus destroying its potential role as DM candidate. 
  More dangerously, in the latter case, it could give rise to the overclosure  of the universe. Since its production mechanism could be non-thermal  any conclusion should be drawn only after a  detailed  analysis that goes beyond the purposes of this work. In any case, we could  let $S$  acquire  a  small vacuum expectation value $\sim v_W^2/M_S$  and  not impose the $Z_2$ symmetry so that the scalar state would rapidly decay into   SM particles through its mixing with  the SM Higgs boson. 
  
\begin{figure}[t!]
\begin{center}
\includegraphics[width=3.2in]{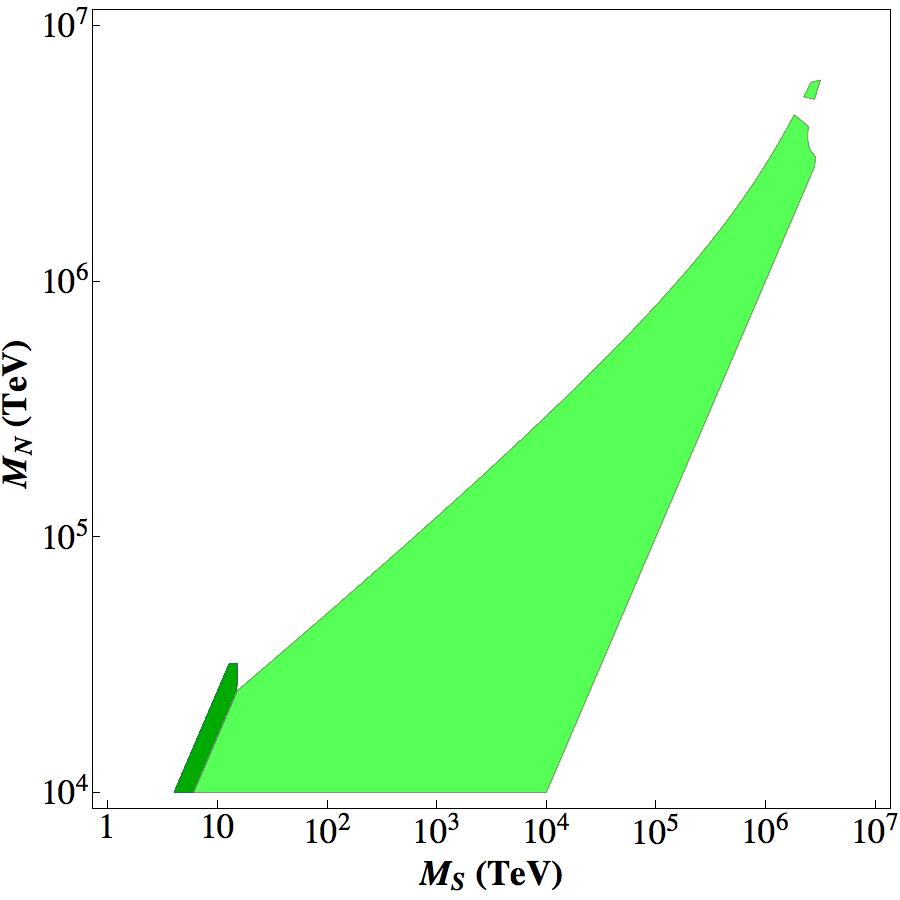}
\caption{\small   See-saw model: The two regions  for which the one-loop contribution vanishes and $\lambda_3$ satisfy \eq{lambda3} (narrow dark green region) and $|\lambda_3|<0.44 M_S/\mbox{TeV}$  (light green region) respectively.  The points have been selected by requiring that: $\lambda_3>-1.6$  to avoid too a large a value for $\lambda_2$, according to \eq{bpot}, the order of $\sqrt{\lambda_3} M_S\lesssim 10$ TeV to  control the two-loop corrections, $M_N\geq M_S$ according to our assumption and $\sum m_nu$ in the range $0.055-0.44$ eV (see the text).
For the points in the narrow dark-green region, the model provide a viable DM candidate, whereas for those in the  light-green region a detailed analysis of the singlet production mechanism should be done before ruling out the model, as commented in the text. 
\label{fig0}}
\end{center}
\end{figure}

In the second model we discussed,   $\lambda_3$ depends only on the ratio $M_Q/M_S$ and it is scale independent, thus the correct relic density may be accommodated for any value of $M_S$, in particular for  $M_S\geq 10$ TeV.

\begin{figure}[t!]
\begin{center}
\includegraphics[width=3.2in]{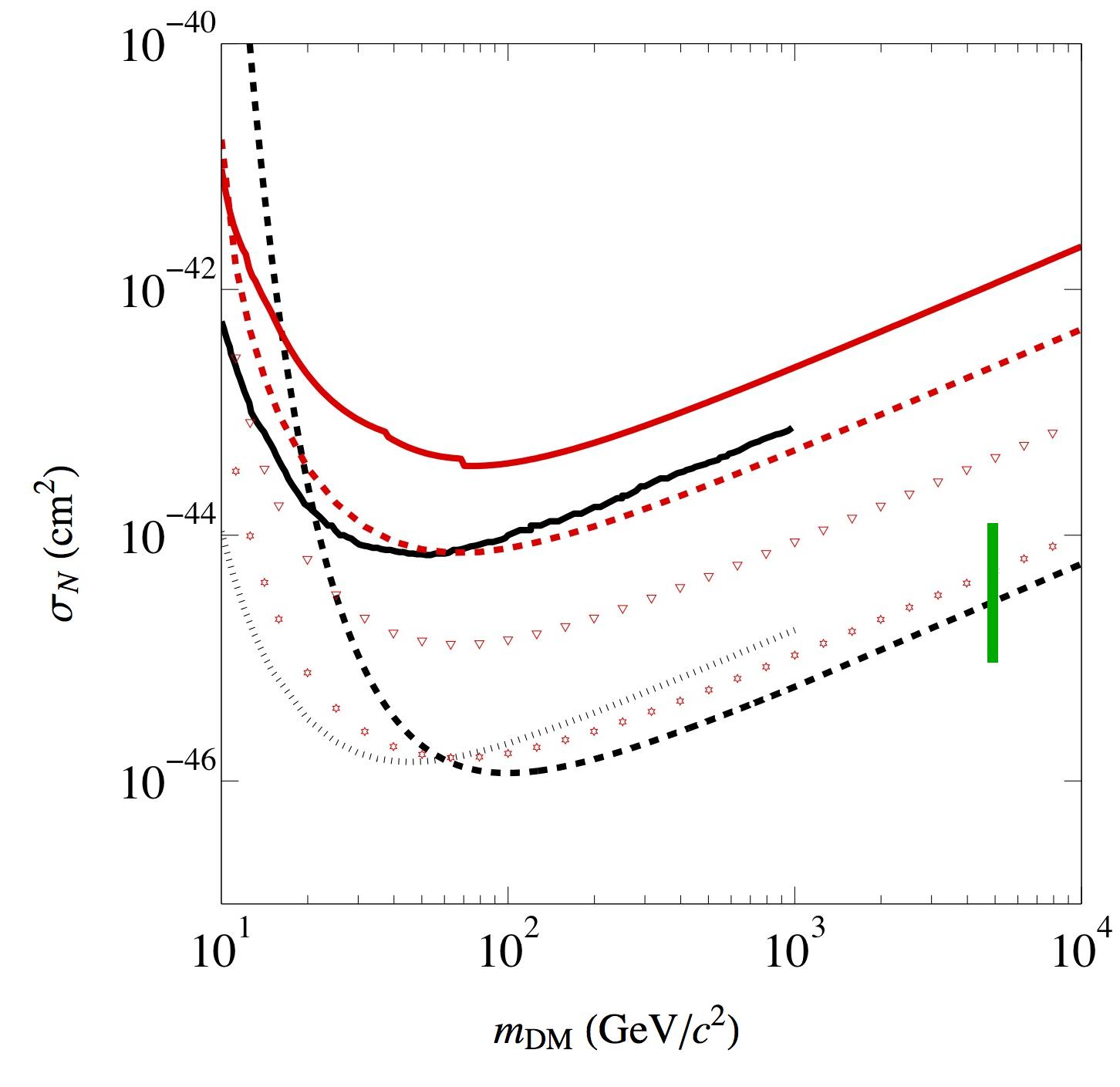}
\caption{\small   Spin independent cross section per nucleon versus  DM candidate masses~\cite{DMT}. The black (red) solid line corresponds to the  XENON100 (CDMSII) data. Black points and the black dashed  line are   the projections for upgraded XENON100  and XENON1T, respectively.  The red dashed line, down triangles and stars correspond to different projections for SCDMS. The green vertical line  is the prediction of the   inert model discussed in this work.
\label{fig}}
\end{center}
\end{figure}

\vskip 0.5cm
{\bf 5.} Let us  briefly  comment on the possibility 
 of detecting the inert scalar $S$  in nuclear scattering experiments. 
 
 The $\lambda_3$ quartic term in \eq{pot} gives rise also to the three fields interaction $SSh$ which yields the effective singlet-nucleon vertex
\be
f_N  \frac{\lambda_3 m_N}{m_h^2} S S \,\bar{\psi}_N \psi_N\,.
\ee
The (non-relativistic) cross section for the process  is given by~\cite{sigma}
\be
\sigma_N = f_N^2 m_N^2 \frac{\lambda_3^2}{4 \pi} \left( \frac{m_r}{m_{S} m^2_h} \right)^2 \,,
\ee
where $m_r$ is the reduced mass for the system which is, to a vary good approximation  in our case, equal to the nucleon mass $m_N$; the  factor $f_N$ contains many uncertainties due to the computation of the nuclear matrix elements and it can vary  from  0.3 to 0.6~\cite{nucleon}. Substituting the values we have found for our model, we obtain, depending on the choice of parameters within the given uncertainties, a cross section 
$\sigma_N$ between $ 10^{-45}$  and  $10^{-44}\mbox{cm}^2$, a value within reach of the next generation of experiments  (see Fig.~\ref{fig}).

\vskip 0.5cm
{\bf 6.} As the scale of new physics is pushed to around the 10 TeV scale or higher, the stability of the Higgs boson mass against finite one-loop corrections induced by the new states give rise to a little hierarchy problem. Since these new states are beyond the current experimental reach, we can use this problem in an heuristic manner to determine masses and couplings of the new particles. We have shown that for two representative new physics scenarios---namely,  see-saw neutrino mass generation and gauge couplings unification---the addition of an inert scalar state suffices in solving the little hierarchy problem and provides in addition a viable candidate for DM. Such a candidate may well be the only experimentally testable signature of the new physics.


\acknowledgments

 We thank P.\ Ullio  for explaining to us some aspects of  DM physics. MF thanks SISSA for the hospitality.


\end{document}